\documentclass[aps,prb,twocolumn,preprintnumbers,amsmath,amssymb,superscriptaddress,floatfix]{revtex4}
\usepackage{graphicx}
\usepackage{dcolumn}
\usepackage{bm}
\usepackage{amsfonts}
\usepackage{bm}
\usepackage{amsmath}
\begin{document}
\title{Chiral decomposition in the electronic structure of graphene multilayers}
\author{Hongki Min and A.H. MacDonald}
\address{Department of Physics, The University of Texas at Austin, Austin, Texas 78712, USA}

\date{\today}

\begin{abstract}
We show that the low-energy electronic structure of arbitrarily stacked
graphene multilayers with nearest-neighbor interlayer tunneling
consists of chiral pseudospin doublets.  Although the number of doublets in
an $N$-layer system depends on the stacking sequence, the pseudospin chirality sum is always $N$.
$N$-layer stacks have $N$ distinct Landau levels at $E=0$ for each spin and valley, and quantized
Hall conductivity $\sigma_{xy} = \pm(4 e^2/h)(N/2+n)$ where $n$ is a non-negative integer.
\end{abstract}

\maketitle
\normalsize
\section{Introduction} 
The recent explosion\cite{reviews} of research on the electronic properties of single layer and stacked multilayer graphene sheets
has been driven by advances in material preparation methods\cite{novoselov2004,berger2004}, by the
unusual\cite{ohta2006,Rycerz2007,Cheianov2007} electronic properties of these
materials including unusual quantum Hall effects\cite{novoselov2005,zhang2005},
and by hopes that these elegantly tunable systems might be useful electronic materials.
In this paper, we demonstrate an unanticipated low-energy property of
graphene multilayers, which follows from an interplay between interlayer tunneling and the chiral
properties of low-energy quasiparticles in an isolated graphene sheet.  Our conclusions
apply in the strongest form to models with only
nearest-neighbor interlayer tunneling, but are valid over a broad field range 
as we explain below.  We find that the low-energy band structure of any graphene multilayer
consists of a set of independent pseudospin doublets.  Within each doublet,
the bands are described by a pseudospin Hamiltonian of the form
\begin{equation}
{\cal H}_J({\bm k}) \, \propto  k^J \; [\, \cos(J \phi_{\bm k}) \, \tau^x \, \pm \, \sin(J \phi_{\bm k}) \, \tau^y \,],
\label{eq:chiralband}
\end{equation}
where $\tau^{\alpha}$ is a Pauli matrix acting on the doublet pseudospin, ${\bm k}$ is an
envelope function momentum measured from either the $K$ or $K'$ corner of the
honeycomb lattice's Brillouin-zone\cite{reviews}, $k=|{\bm k}|$, and $\phi_{\bm k}$ is the
orientation of ${\bm k}$.  The $\pm$ sign in Eq. (\ref{eq:chiralband}) assumes the opposite signs in
graphene's $K$ and $K'$ valleys.  Following the
earlier work on graphene bilayers\cite{mccann2006}, we refer to $J$ as the chirality
index of a doublet.  In the presence of a perpendicular magnetic field $B$, ${\cal H}_J({\bm k})$ yields the $J$ Landau levels at $E = 0$ and $E \ne 0$ levels with $|E| \propto B^{J/2}$.  Taking the twofold spin and valley degeneracies into account,
the number of independent zero-energy band eigenstates at the Dirac
point (${\bm k}=0$) is therefore $8 N_{D}$, where $N_D$ is the number of pseudospin doublets.
We find that, although $N_{D}$ depends on the details of the stacking sequence,
\begin{equation}
\sum_{i=1}^{N_{D}} \; J_i \; = N
\label{eq:sumrule}
\end{equation}
in an $N$-layer stack.
It follows from
Eq. (\ref{eq:sumrule}) that the Hall conductivity of an
$N$-layer stack has strong integer quantum Hall effects with plateau conductivities,
\begin{equation}
\sigma_{xy} \, = \,\pm {4 e^2\over h} \,\left({N\over 2}+n\right),
\label{eq:iqhe}
\end{equation}
where $n$ is a non-negative integer.

\begin{figure}[h]
\includegraphics[width=0.8\linewidth]{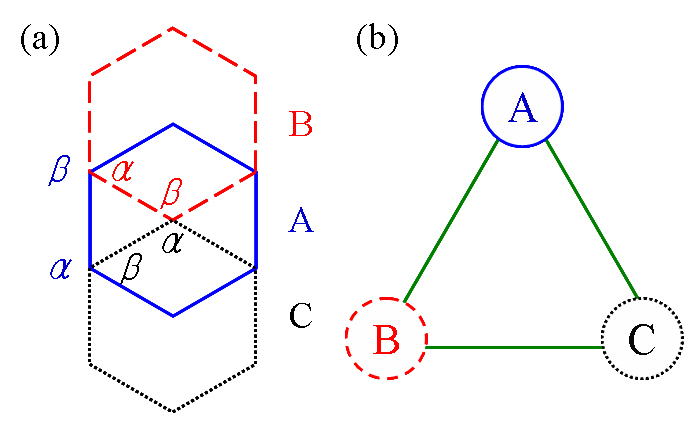}
\caption{(Color online) (a) Energetically favored stacking arrangements for graphene sheets.
The honeycomb lattice of a single sheet has two triangular sublattices, labeled by
$\alpha$ and $\beta$.  Given a starting graphene sheet, the honeycomb lattice for the next layer is positioned
by displacing either $\alpha$ or $\beta$ sublattice carbon atoms along a honeycomb edge.
This results in three distinct two-dimensional sheets, labeled by A, B, and C.
Representative $\alpha$ and $\beta$ sublattice positions in A, B, and C layers are identified in this illustration.
It is also possible to transform between layer types by rotating by 60$^\circ$ about one
of the carbon atoms.  (b) Each added layer cycles around this stacking triangle in either the
right-handed or the left-handed sense.  Reversals of the sense of this rotation tend
to increase the number of low-energy pseudospin doublets $N_D$.  In graphite, the Bernal (AB) stacking
corresponds to a reversal at every step and orthorhombic (ABC) stacking corresponds to no reversals.
}
\label{fig:stack}
\end{figure}

\section{Stacking Diagrams and Partitioning Rules}
When one graphene layer is placed on another, it is energetically favorable\cite{charlier1992}
for the atoms of either $\alpha$ or $\beta$ sublattices to be displaced along the honeycomb edges,
as illustrated in Fig. \ref{fig:stack}.  This stacking rule implies the three distinct but equivalent projections
(labeled A, B, and C) of the three-dimensional structure's honeycomb-lattice layers onto the $\hat{x}$-$\hat{y}$ plane
and $2^{N-2}$ distinct $N$-layer stack sequences.  When a B layer is placed on an A layer, a C layer on a B layer, or an A layer on a C layer,
the $\alpha$ sites of the upper layer are above the $\beta$ sites of the lower layer and therefore 
linked by the nearest interlayer neighbor $\pi$-orbital hopping amplitude $t_{\perp}$.
For the corresponding anticyclic stacking choices (A on B, B on C, or C on A), it is the $\beta$ sites of the upper layer and the
$\alpha$ sites of the lower layer that are linked.

\begin{figure}[h]
\includegraphics[width=1\linewidth]{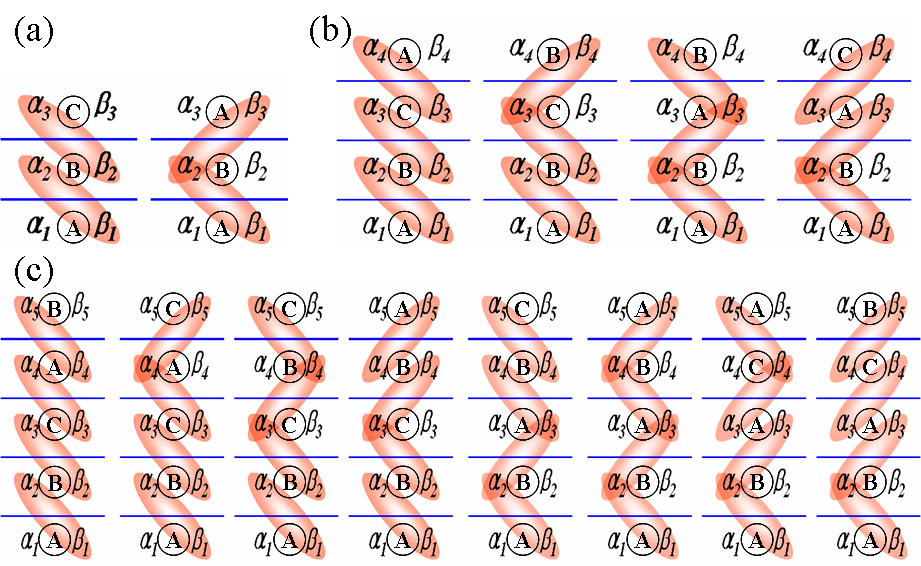}
\caption{(Color online)  Stacking sequences and linkage diagrams for $N=3,4,5$ layer stacks.
The low-energy band and the Landau level structure of a graphene stack are readily read off these diagrams
as explained in the text.  Shaded ovals link
$\alpha$ and $\beta$ nearest interlayer neighbors.}
\label{fig:min_diagrams}
\end{figure}

All distinct $N=3$, $N=4$, and $N=5$ layer
stacks are illustrated in Fig. \ref{fig:min_diagrams}, in which we have arbitrarily labeled the first two layers starting
from the bottom as A and B.  The low-energy band and the Landau level structure can be read off these diagrams
by partitioning a stack using the following rules,
which are justified in the following section.
(i) Identify the longest nonoverlapping segments within which there
are no reversals of stacking sense.  When there is ambiguity in the selection of nonoverlapping segments,
choose the partitioning which incorporates the largest number of layers.  Each segment (including for
interior segments the end layers
at which reversals take place) defines a $J$-layer partition of the stack and may be associated with
a chirality $J$ doublet.  (ii)  Iteratively partition of the remaining segments of the stack into
smaller $J$ elements, excluding layers
contained within previously identified partitions, until all layers are exhausted.
Chirality decompositions which
follow from these rules are summarized in Table \ref{tab:decomposition}.
Note that this procedure can result in $J=1$ doublets associated with separated 
single layers which remain at the last step in the partitioning process.

\begin{figure}[h]
\includegraphics[width=0.7\linewidth]{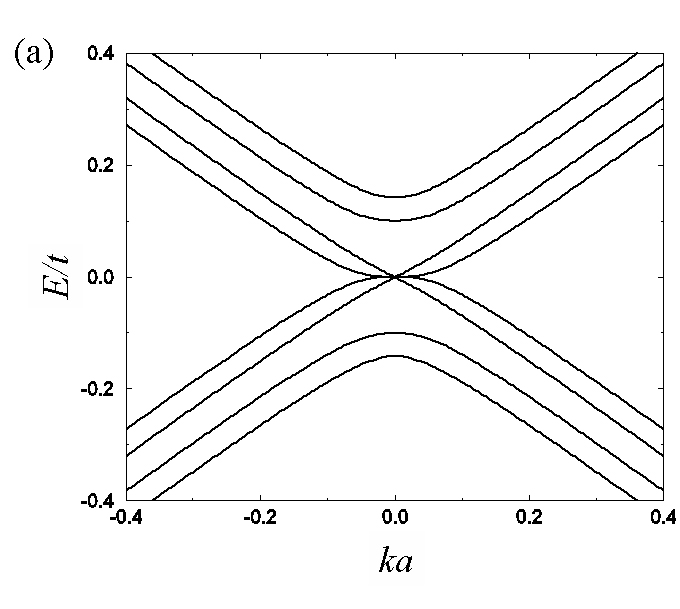}
\includegraphics[width=0.7\linewidth]{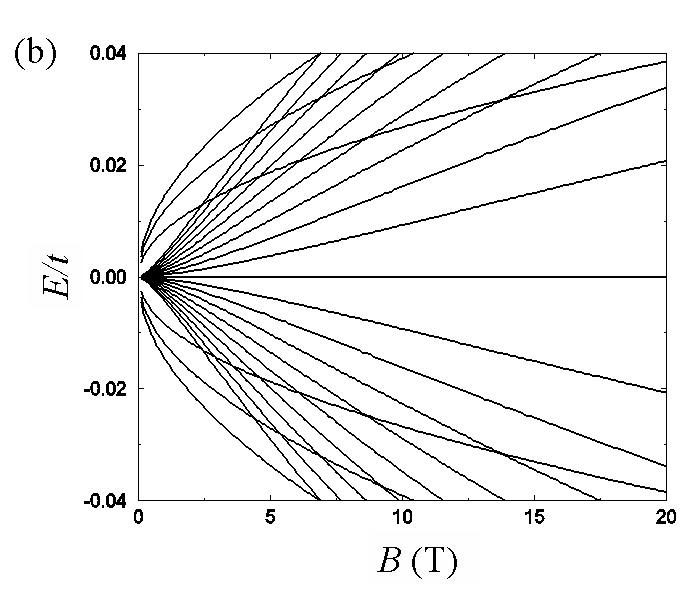}
\caption{(a) Band structure and (b) low-energy Landau levels
of tetralayer graphene with ABCB stacking.
These bands were evaluated for nearest intralayer neighbor hopping $t=3$ eV and nearest interlayer neighbor hopping $t_{\perp}=0.1t$.}
\label{fig:stack_ABCB}
\end{figure}

In applying these rules, the simplest case is cyclic ABC stacking for which there are
no stacking sense reversals and therefore a single $J=N$ partition.
In the opposite limit, AB stacking, the stacking sense is reversed in every layer and the rules
imply $N/2$ partitions with $J=2$ for even $N$, and when $N$ is odd a remaining $J=1$ partition.  Between these two limits, a rich variety of qualitatively distinct low-energy behaviors occur.
For example, in the ABCB stacked tetralayer, ABC is identified as a $J=3$ doublet and the remaining B
layer gives a $J=1$ doublet.  The low-energy band structure and the Landau level structure
of this stack, as illustrated in Fig. \ref{fig:stack_ABCB}, have
two sets of low-energy bands with $|E|\propto k, k^3$,
Landau levels with $|E|\propto B^{1/2},B^{3/2}$, and four zero-energy Landau levels
per spin and valley.  All these properties are predicted by the partitioning rules.
We have explicitly checked that the rules correctly reproduce the low-energy electronic structure
for all stacking sequences up to $N=7$.
Because each layer is a member of one and only one partition, the partitioning rules imply
the chirality sum rule in Eq. (\ref{eq:sumrule}).

\begin{table}[h]
\begin{ruledtabular}
\begin{tabular}{p{0.12\textwidth} p{0.12\textwidth} | p{0.12\textwidth} p{0.12\textwidth}}
stacking & chirality & stacking & chirality\\
\hline
ABC      & 3         & ABCABC   & 6        \\
ABA      & 2+1       & ABCABA   & 5+1      \\
         &           & ABCACA   & 4+2      \\
ABCA     & 4         & ABCACB   & 4+2      \\
ABCB     & 3+1       & ABCBCA   & 3+3      \\
ABAB     & 2+2       & ABCBCB   & 3+2+1    \\
ABAC     & 1+3       & ABCBAB   & 3+2+1    \\
         &           & ABCBAC   & 3+3      \\
ABCAB    & 5         & ABABCA   & 2+4      \\
ABCAC    & 4+1       & ABABCB   & 2+3+1    \\
ABCBC    & 3+2       & ABABAB   & 2+2+2    \\
ABCBA    & 3+2       & ABABAC   & 2+1+3    \\
ABABC    & 2+3       & ABACAB   & 2+1+3    \\
ABABA    & 2+2+1     & ABACAC   & 1+3+2    \\
ABACA    & 1+3+1     & ABACBC   & 1+4+1    \\
ABACB    & 1+4       & ABACBA   & 1+5      \\ 
\end{tabular}
\caption{Chirality decomposition for $N=3,4,5,6$ layer stacks.}
\label{tab:decomposition}
\end{ruledtabular}
\end{table}

\section{Degenerate State Perturbation Theory}
We start from the well-known $J=1$ massless
Dirac equation\cite{reviews} ${\bm k}\cdot{\bm p}$ model for isolated sheets,
\begin{equation}
\label{eq:hamiltonian_MD}
{\cal H}_{MD}({\bm k})=-\left(
\begin{array}{cc}
0              &v\pi^{\dagger}   \\
v\pi           &0                \\
\end{array}
\right),
\end{equation}
where $\pi = \hbar v(k_x + i k_y)$
and $v$ is the quasiparticle velocity.
In the presence of an external magnetic field,
$\pi^{\dagger}$ and $\pi$ are proportional to the Landau level raising and lowering operators, so that Eq. (\ref{eq:hamiltonian_MD})
implies the presence of one macroscopically degenerate Landau level at the Dirac point for each spin and valley, and therefore,
to the $N=1$ quantum Hall effect\cite{novoselov2005,zhang2005} of Eq. (\ref{eq:iqhe}).
An $N$-layer stack has a two-dimensional band structure with $2N$ atoms per unit cell.
The Hamiltonian can be written as 
\begin{equation}
{\cal H} = {\cal H}_{\perp} + {\cal H}_{\parallel},
\end{equation}
where ${\cal H}_{\perp}$ accounts for interlayer tunneling and ${\cal H}_{\parallel}$ for intralayer tunneling.  ${\cal H}_{\parallel}$ is
the direct product of massless Dirac model Hamiltonians ${\cal H}_{MD}$ for the sublattice pseudospin degrees of freedom of each layer.  We construct a low-energy Hamiltonian by first identifying the zero-energy eigenstates of $H_{\perp}$
and then treating ${\cal H}_{\parallel}$ as a perturbation.

Referring to Fig. \ref{fig:min_diagrams}, we see that
${\cal H}_{\perp}$ is the direct product of a set of finite-length one-dimensional (1D) tight-binding chains
and a null matrix with dimension equal to the
number of isolated sites. An elementary calculation gives the
following eigenvalues and eigenvectors for a chain of length $M$:
\begin{eqnarray}
\varepsilon_r&=&2 \, t_{\perp} \, \cos\theta_r, \\
{\bm a}_r &=& \sqrt{2\over M+1}(\sin\theta_r,\sin 2\theta_r,\cdots,\sin M\theta_r), \nonumber
\label{eq:chain}
\end{eqnarray}
where $r = 1,2,\ldots, M$ is the chain eigenvalue index and
$\theta_r=r\pi/(M+1)$.  Note that odd $M$ chains have a zero-energy eigenstate with an
eigenvector that has nonzero amplitudes, constant in magnitude and alternating in sign,
on the sublattice of the chain ends.
The set of zero-energy eigenstates of ${\cal H}_{\perp}$ consists of the states
localized on isolated sites and the single zero-energy eigenstates of each odd-length chain.

The low-energy effective Hamiltonian is evaluated by applying leading order
degenerate state perturbation theory to the zero-energy subspace.
The matrix element of the effective Hamiltonian between degenerate zero-energy states $r$ and $r'$
is given by\cite{sakurai1994}
\begin{equation}
\label{eq:perturbation}
\left<\Psi_r|{\cal H}|\Psi_{r'}\right>= \left<\Psi_r\right| {\cal H}_{\parallel}
\left[ \hat{Q}(-H_{\perp}^{-1}) \hat{Q} {\cal H}_{\parallel} \right]^{n-1}\left|\Psi_{r'}\right>,
\end{equation}
where $n$ is the smallest positive integer for which the matrix element is nonzero,
and $\hat{Q}=1-\hat{P}$, $\hat{P}$ is a projection operator onto the zero-energy subspace.  To understand the
structure of this Hamiltonian, it is helpful to start with some simple examples.


For ABC stacked $N$-layer graphene, the zero-energy states are the two isolated site states in top and
bottom layers, $\alpha_1$ and $\beta_N$.
The high-energy Hilbert space is formed from a set of two-site chains.
Because $H_{\parallel}$ is diagonal in layer index and
$H_{\perp}$ (and hence $H_{\perp}^{-1}$) can change the layer index by one unit,
the lowest order at which $\alpha_1$ and $\beta_N$ are coupled is $n=N$.
It follows that
\begin{equation}
\label{eq:ABC_eff}
\left<\alpha_1|{\cal H}|\beta_N\right>
= \; -t_{\perp}\left(v\pi^{\dagger}/t_{\perp}\right)^N.
\end{equation}
Thus, the effective Hamiltonian of $N$-layer graphene with ABC stacking has a
single $J=N$ doublet\cite{previouswork}.

For AB stacked $N$-layer graphene, the high-energy Hilbert space
consists of a single $N$-site 1D chain, excluding its zero- energy eigenvalue when $N$ is odd.  
There is an isolated site in each layer which is connected to both its neighbors at order
$n=2$ forming an isolated site chain.  When $N$ is even, this chain
is diagonalized by $N/2$, $J=2$ doublets formed between $\alpha$-sublattice and $\beta$-sublattice
chain states\cite{previouswork}.  When $N$ is odd, the zero-energy chain state is mapped to
an equal-magnitude oscillating-sign linear combination of isolated site states by intralayer tunneling
at order $n=1$, yielding a $J=1$ doublet. The $(N-1)/2$, $J=2$ doublets are then
formed between $\alpha$-sublattice and $\beta$-sublattice isolated site chain states
in the orthogonal portion of the isolated state subspace.

A more complex and more typical example is realized by
placing a single reversed layer on top of ABC stacked N-layer graphene with $N>2$.
In this case, we obtain
\begin{equation}
\label{eq:hamiltonian_ABCB_eff}
{\cal H}_{N+1}^{eff}=-t_{\perp}\left(
\begin{array}{cccc}
0                  & {\nu^{\dagger}\over \sqrt{2}}          & 0                            & {(\nu^{\dagger})^2\over 2}  \\
{\nu\over\sqrt{2}} & 0                                      & -{(\nu)^{N-1}\over \sqrt{2}} & 0                           \\
0                  & -{(\nu^{\dagger})^{N-1}\over \sqrt{2}} & 0                            & {(\nu^{\dagger})^N\over 2}  \\
{\nu^2\over 2}     & 0                                      & {(\nu)^N\over 2}             & 0                           \\
\end{array}
\right),
\end{equation}
using a $(\alpha_{N+1},\beta_{N+1}^{-},\alpha_1,\beta_N)$ basis.
Here, $\nu=v\pi/t_{\perp}$ and $\beta_{N+1}^{-}$ is
the three-site chain zero-energy state, $\beta_{N+1}^{-}=\left(\beta_{N+1}-\beta_{N-1}\right)/\sqrt{2}$.
The first $2\times 2$ block in Eq. (\ref{eq:hamiltonian_ABCB_eff}) gives a $J=1$ doublet with a reduced velocity.
The $J=N$ doublet in this instance includes both the $(\alpha_1,\beta_N)$ subspace contribution and
an equal contribution due to perturbative coupling to the $(\alpha_{N+1},\beta_{N+1}^{-})$ subspace.
The final Hamiltonian is reduced to
\begin{equation}
{\cal H}_{N+1}^{eff}\approx{\cal H}_1\otimes {\cal H}_N,
\end{equation}
where
\begin{equation}
{\cal H}_1=-t_{\perp}\left(
\begin{array}{cc}
0            & \nu^{\dagger}/\sqrt{2} \\
\nu/\sqrt{2} & 0                      \\
\end{array}
\right), \ \
{\cal H}_N=-t_{\perp}\left(
\begin{array}{cc}
0       & (\nu^{\dagger})^N \\
(\nu)^N & 0                 \\
\end{array}
\right).
\end{equation}

The relationship between the electronic structure of a general stack and
the partitioning procedure explained above can be understood as follows.  First, note that a partition with chirality $J$ has isolated sites in its terminal layers that are coupled at order $J$ in perturbation theory.
In the case of $J=1$ partition, the chain opposite to the
single isolated site always has an odd length and provides the zero-energy partner;
isolated site to chain coupling therefore always occurs at first order.
Next, consider the perturbation theory, truncating at successively higher orders.
When truncated at first order, the $J=1$ partitions are isolated
by higher $J$ blocks within which the Hamiltonian vanishes.  Each $J=1$
partition therefore yields a separate massless Dirac equation
with velocities\cite{velocity} that can be smaller than the graphene sheet
Dirac velocity.  When the perturbation theory is truncated at second order,
the Hamiltonian becomes nonzero within the $J=2$ partitions.  The eigenenergies
within the $J=1$ partitions are parametrically larger, and the
Hamiltonian within the $J>2$ partitions is still zero.  To leading order therefore,
the $J=2$ partitions are separated, and their isolated states are coupled at
the second order in perturbation theory so that each provides a $J=2$ doublet such as that of an
isolated bilayer.  If two or more $J=2$ partitions are adjacent, then their
Hamiltonians do not separate.  In this case, there is a chain of second order
couplings between isolated states, such as those of an even-length AB stack, but
the end result is still $J=2$ doublet for each $J=2$ partition.
The identification between partitions and chiral doublets can be established by
continuing this consideration up to the highest values of $J$ which occur for a particular
stack. 
Then, the effective Hamiltonian of any $N$-layer graphene is as follows: 
\begin{equation}
{\cal H}_N^{eff}\approx{\cal H}_{J_1} \otimes {\cal H}_{J_2} \otimes \cdots \otimes {\cal H}_{J_{N_D}},
\end{equation}
with the sum rule in Eq. (\ref{eq:sumrule}).
Note that $N_D$ is half the sum of the number of isolated sites and the number of odd-length chains.

\section{Discussion}
\begin{figure}[h]
\includegraphics[width=1\linewidth]{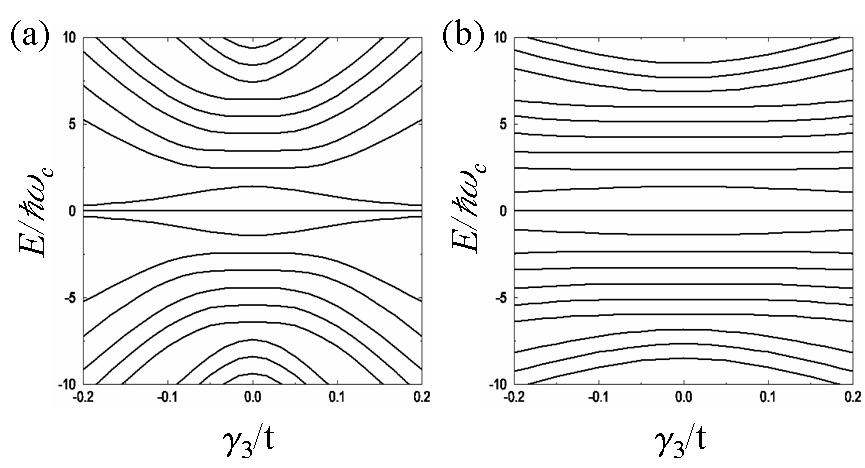}
\includegraphics[width=1\linewidth]{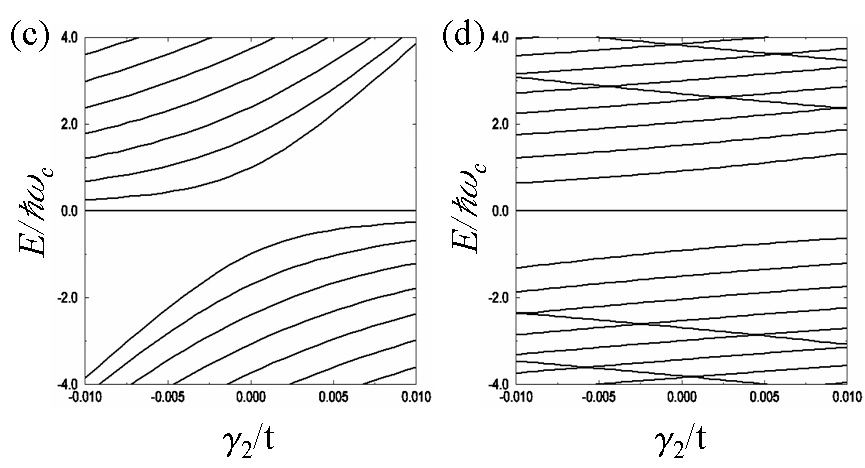}
\caption{Landau level spectrum at the $K$ valley as a function of $\gamma_3$ for an AB stacked bilayer for (a) $B=0.1$ T and (b) $B=1$ T, and as a function of $\gamma_2$ for an ABA stacked trilayer for (c) $B=1$ T and (d) $B=10$ T. Here $t=3$ eV, $t_{\perp}=0.1t$, and $\omega_c=eB/mc$, with $m=t_{\perp}/2v^2$, were used.}
\label{fig:2NN}
\end{figure}

The minimal model we have used to derive these results is approximately 
valid in the broad intermediate magnetic field $B$ range between $\sim 10$ and $\sim 100$ T,
over which the intralayer hopping energy in field ($\sim \hbar v/\ell$ where $\ell = \sqrt{\hbar c/eB} \sim 25\, {\rm nm}/[B({\rm T})]^{1/2}$ is the magnetic length) is larger than the distant neighbor interlayer 
hopping amplitudes that we have neglected ($\gamma_2 \sim -20 $ meV), 
but still smaller than $t_{\perp}$.
For example, if we consider $\alpha_1\rightarrow \alpha_3$ hopping process in ABA stacked trilayer in Fig. \ref{fig:min_diagrams}, the valid range of magnetic field for the minimal model is given by 
\begin{equation}
|\gamma_2| < {(\hbar v/l)^2 \over t_{\perp}} < t_{\perp}.
\end{equation}

When $\gamma_2$ does not play an important role (in $N=2$ stacks, for example), the lower limit of the validity range 
is parametrically smaller.  The minimum field in bilayers has been estimated to be $\sim 1$ T\cite{mccann2006}, by comparing intralayer hopping with the $\gamma_3 \sim 0.3$ eV interlayer hopping amplitude,
\begin{equation}
\hbar v_3/l  < {(\hbar v/l)^2 \over t_{\perp}} < t_{\perp},
\end{equation}
where $v_3=(\sqrt{3}/2)a\gamma_3/\hbar$ and $a$ is a lattice constant of graphene.

Figure \ref{fig:2NN} shows the Landau level spectrum at the $K$ valley as a function of $\gamma_3$ for an AB stacked bilayer and as a function of $\gamma_2$ for an ABA stacked trilayer. In the case of the bilayer, the dependence of the Landau levels on $\gamma_3$ is weak 
for $B$ larger than 1 T, whereas in the case of the trilayer, the Landau level spectrum still strongly depends on $\gamma_2$ for $B=1$ T, but the dependence becomes weak for $B$ above 10 T, confirming the above argument.

\begin{figure}[h]
\includegraphics[width=0.8\linewidth]{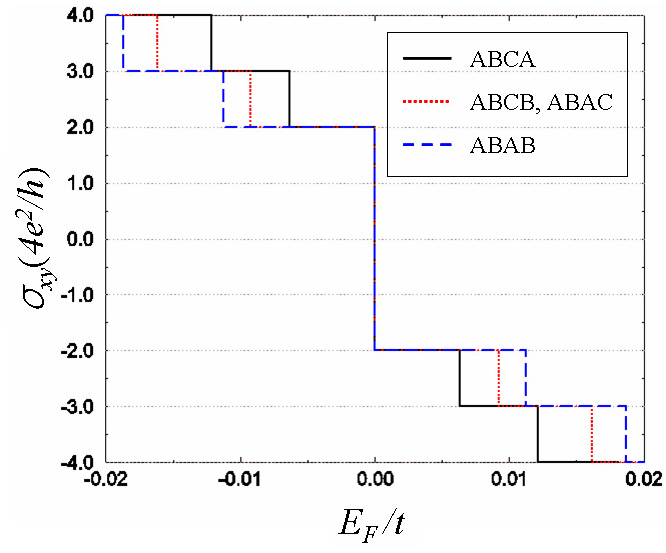}
\caption{(Color online) Noninteracting Hall conductivity as a function of the Fermi energy
for all inequivalent four-layer graphene stacks when $B=10$ T, $t=3$ eV, and $t_{\perp}=0.1t$.}
\label{fig:IQHE4}
\end{figure}

In Fig. \ref{fig:IQHE4}, we plot the noninteracting Hall conductivity as a
function of the Fermi energy for the four distinct four-layer stacks.
When electron-electron interactions are included at an electrostatic mean-field (Hartree) level, 
charge redistribution changes the positions of plateaus which are away from the Dirac point, but it does not alter 
the conductivity jumps at the Dirac point. If we allow the remote hopping, a small $\sigma_{xy}=0$ plateau appears 
at the Dirac point for the ABAB staked tetralayer due to the energy gap opened by $\gamma_2$.  This small gap 
is likely to be important only in very weak disorder samples, in which case electron-electron
interactions beyond the Hartree level are also likely to be important\cite{refs_qhf}. 
The property that the Hall conductivity jumps by four units on crossing the Dirac point for arbitrarily stacked tetralayer 
graphene is the most obvious experimental manifestation of the chirality sum rule discussed in this paper.
In practice charged multilayers ($E_F\neq 0$) would normally be prepared by placing the system on one side of an electrode and gating.
Even though gating will redistribute charge and shift energies differently in different layers, the Landau level bunching we discussed
should still be clearly reflected in quantum Hall effect measurements.

In numerical calculations, we have found that 
for many stacks, the $N$-fold degeneracy at $E=0$ persists even outside the field range
over which the minimal model is accurate.
Because of the topological character of the quantum Hall effect, we
expect the plateaus at $|\sigma_{xy}| = (4 e^2/h) N/2$ to be
exceptionally robust in most stacks, and that plateaus at the smaller 
Hall conductivity samples will require exceptionally high quality samples.  
We do not anticipate strong quantum Hall effects with $|\sigma_{xy}| < (4 e^2/h) N/2$  
unless gaps between the Landau levels are enhanced by interactions\cite{refs_qhf}
in very high quality\cite{nomura2006} samples.  



Finally, we note that chiral two-dimensional electron system tend toward momentum-space vortex states
in which charge is spontaneously shifted between layers\cite{min2007}
and that these instabilities are stronger in systems with larger $J$.
The present work identifies ABC stacked multilayer graphene as the most
likely candidate for these exotic states.

\acknowledgments
This work was supported by the Welch Foundation, by NSF-NRI SWAN, and by the National Science Foundation
under grant DMR-0606489.



\begin{thebibliography}{999}
\bibitem{velocity}  The velocity of the $J=1$ doublets is determined by the strength of the coupling
between the odd-length chain zero-energy state and isolated states on the sublattice opposite to the chain ends.
For a chain of length $2N-1$, the chain's zero-energy state has nonzero amplitude on the $N$ odd-index sites.
The velocity is reduced from the single sheet velocity by a factor of $\sqrt{M/N}$, where $M$ is the number of
isolated sites opposite to the $N$ odd-index sites.  In a similar manner, higher $J$ doublet Hamiltonians are sometimes altered by a multiplicative factor by perturbative coupling to smaller $J$ doublets
as in the single reversed layer example.
\end{thebibliography}
\end{document}